\documentclass[prl,aps,singlecolumn,superscriptaddress,preprintnumbers,showpacs,tightenlines,floatfix,amsmath,amssymb]{revtex4}

\usepackage{bm}

\usepackage{amsmath}
\usepackage{graphicx}
\usepackage{amssymb}
\usepackage{mathrsfs}
\usepackage{bbm}

\newcommand{\be}{\begin{eqnarray}}
\newcommand{\ee}{\end{eqnarray}}

\begin{document}

\hskip .0cm
\title{{\large \bf On Lorentz Invariance, Spin-Charge Separation And SU(2) Yang-Mills Theory }}

\author{Antti J. Niemi}
\email{Antti.Niemi@physics.uu.se}
\affiliation{Department of Physics and Astronomy, Uppsala University,
P.O. Box 803, S-75108, Uppsala, Sweden}
\affiliation{
Laboratoire de Mathematiques et Physique Theorique
CNRS UMR 6083, F\'ed\'eration Denis Poisson, Universit\'e de Tours,
Parc de Grandmont, F37200, Tours, France}
\affiliation{Chern Institute of Mathematics, Tianjin 300071, P.R. China}
\author{Sergey Slizovskiy}
\email{Sergey.Slizovskiy@fysast.uu.se}
\affiliation{Department of Physics and Astronomy, Uppsala University,
P.O. Box 803, S-75108, Uppsala, Sweden}

\preprint{UUITP-05/09}

\begin{abstract}
Previously it has been shown that in spin-charge separated
SU(2) Yang-Mills theory Lorentz invariance can become
broken by a one-cocycle that appears in the Lorentz 
boosts. Here we study in detail the structure of 
this one-cocycle. In particular we show that 
its non-triviality relates to the presence of a
(Dirac) magnetic monopole bundle. We also explicitely
present the finite version of the cocycle.
\end{abstract}


\date{\today}

\maketitle

Recently  the properties of 
four dimensional SU(2) Yang-Mills theory have been investigated
using spin-charge separated variables \cite{fad1}, \cite{fad2} that might 
describe the confining strong coupling regime of the theory \cite{walet}. 
For example, it was shown that even though these variables
reveal the presence of two massless Goldstone modes,
this apparent contradiction with the existence of a mass 
gap becomes resolved since these Goldstone modes 
break Lorentz invariance by a one-cocycle \cite{fad1}: The
ground state must be Lorentz invariant, thus the one-cocycle 
is to be removed. This demand fixes the ground state
uniquely and deletes all massless states 
from the spectrum \cite{fad1}.

In \cite{fad1} only the infinitesimal form of the 
one-cocycle was presented. Here we display its finite
form. We also verify that the one-cocycle is 
indeed non-trivial, by relating it to the nontriviality of 
the Dirac magnetic monopole bundle.
For definiteness we develop our arguments in the four dimensional 
space $\mathbb R^4$ with Euclidean signature. The extension from
$SO(4)$ to $SO(3,1)$ is straightforward.

Locally, in the Maximal Abelian Gauge the spin-charge 
separation amounts to the following decomposition of the 
off-diagonal components $A^\pm_\mu$ of the gauge 
field $A^a_\mu$ \cite{fad1}, \cite{walet}.
\begin{equation}
A^+_\mu \ = \ A_\mu^1 + i A^2_\mu \ = \ 
\psi_1 e_\mu + \psi_2 e^\star_\mu
\label{A}
\end{equation}
where the spin field $e_\mu$
\[
e_\mu = \frac{1}{\sqrt{2}} ( e^1_\mu + i e^2_\mu )
\]
is normalized according to
\begin{equation}
\begin{matrix}
e_\mu e_\mu = 0 \\
e_\mu e_\mu^\star = 1
\end{matrix}
\label{enorm}
\end{equation}
This can be viewed as a Clebsch-Gordan type decomposition
of $A^\pm_\mu$, when interpreted as a tensor product of 
the complex spin-variable $e_\mu$ that
remain intact under SU(2) gauge transformations
and the charge variables $\psi_{1,2}$ that are Lorentz
scalars but transform under SU(2); see \cite{fad1} for details. 

The decomposition introduces an internal $U_I(1) \times \mathbb Z_2$
symmetry that is not visible to $A^a_\mu$. The $U_I(1)$ action is
\begin{equation}
U_I(1): \ \ \ \ \ \begin{matrix}
e^{}_\mu & \to & e^{-i \lambda} e^{}_\mu \\
\psi^{}_1 & \to & e^{i\lambda}\psi^{}_1 \\
\psi^{}_2 & \to & e^{-i\lambda}\psi^{}_2
\end{matrix}
\label{intu1}
\end{equation} 
This is a local frame rotation, in particular it preserves the
orientation in $e^{}_\mu$. The $\mathbb Z_2$ action 
exchanges $\psi_1$ and $\psi_2$, 
\begin{equation}
\mathbb Z_2: \ \ \ \ \ \ \begin{matrix}
e^{}_\mu & \to & e^{\star}_\mu \\
\psi^{}_1 & \to & \psi^{}_2 \\
\psi^{}_2 & \to & \psi^{}_1
\end{matrix}
\label{intz2}
\end{equation} 
This changes the orientation on the two-plane spanned by $e_\mu$.
(The realization of $\mathbb Z_2$ is unique only
up to phase factor.)

In the Yang-Mills action the complex scalar fields 
$\psi_{1,2}$ becomes combined into the three component 
unit vector \cite{fad1}
\begin{equation}
{\bf t}  \ = \ \frac{1}{\rho^2} 
\left( \begin{matrix} \psi^\star_1 & \psi^\star_2 \end{matrix} \right)
\vec \sigma \left( \begin{matrix} \psi^{}_1 \\ 
\psi^{}_2 \end{matrix} \right)
\ = \ \frac{1}{\rho^2} \left( \begin{matrix} 
\psi^\star_1 \psi^{}_2 + \psi^\star_2 \psi^{}_1 \\
i (\psi^{}_1 \psi^\star_2 - \psi^{}_2 \psi^\star_1) \\
\psi^\star_1 \psi^{}_1 - \psi^{\star}_2 \psi^{}_2 \end{matrix}
\right)
\ = \ 
\left( \begin{matrix} \cos \phi  \sin \theta \\ \sin\phi 
\sin \theta \\ \cos \theta \end{matrix} \right)
\label{n}
\end{equation}
We have here parameterized
\begin{equation}
\begin{matrix}
\psi^{}_1 & = & \rho \, e^{i\zeta} \cos \frac{\theta}{2} \,
e^{-i\phi/2} \\
\psi^{}_2 & = & \rho \, e^{i\zeta} \sin \frac{\theta}{2} \,
e^{i \phi/2}
\end{matrix}
\label{psipara}
\end{equation}
The internal $U_I(1)$ transformation sends
\begin{equation}
t_\pm \ = \ \frac{1}{2} (t_1 \pm i t_2 ) \ \rightarrow \ e^{\mp 2
i \lambda} t_\pm
\label{tpm}
\end{equation}
but $t_3$ remains intact. The 
$\mathbb Z_2$ action is a rotation that sends $(t_1, t_2 , 
t_3) \to (t_1 , -t_2 , -t_3)$.  In terms of the 
angular variables in (\ref{n}) this corresponds
to $(\phi, \theta) \to (2\pi - \phi , \pi - \theta)$.
Thus we may opt to eliminate the $\mathbb Z_2$ degeneracy by 
a restriction to the upper hemisphere $\theta \in [0,\frac{\pi}{2})$.

The off-diagonal components (\ref{A}) determine the embedding of
a two dimensional plane in $\mathbb R^4$. The space
of two dimensional linear subspaces of 
$\mathbb R^4$ is the real Grassmannian manifold $Gr(4,2)$ \cite{grass},
it can be described by the antisymmetric tensor \cite{fad1}, \cite{marsh},
\cite{sergey}
\begin{equation}
P_{\mu\nu} = \frac{i}{2} ( A^+_\mu A^-_\nu - A_\nu^+ A_\mu^-)
\ = \ A^1_\mu A^2_\nu - A^1_\nu A^2_\mu
\label{P}
\end{equation}
that obeys the Pl\"ucker equation
\begin{equation}
P_{12} P_{34} - P_{13}P_{24} + P_{23} P_{14} = 0
\label{PP}
\end{equation}
Conversely,
{\it any} real antisymmetric matrix $P_{\mu\nu}$ that
satisfies (\ref{PP}) can be represented in the functional form
(\ref{P}) in terms of some two vectors $A^1_\mu$ and $A^2_\mu$.
The Pl\"ucker equation describes the embedding of
$Gr(4,2)$ in the five dimensional projective space $\mathbb R \mathbb 
P^5$ as a degree four hypersurface \cite{grass},
a homogeneous space
\begin{equation}
Gr(4,2) \simeq \frac{SO(4)}{SO(2) \times SO(2)} \ \simeq \
\mathbb S^2 \times \mathbb S^2
\label{Gr}
\end{equation}
When we substitute (\ref{A}) we get
\begin{equation}
P^{}_{\mu\nu} = \frac{i}{2}(|\psi_1|^2 - |\psi_2|^2) \cdot 
(e^{}_\mu {e}^{\star}_\nu
- e^{}_\nu {e}^{\star}_\mu) \ = \ \frac{i}{2}\cdot \rho^2 \cdot 
t^{}_3 \cdot (e^{}_\mu {e}^{\star}_\nu
- e^{}_\nu {e}^{\star}_\mu) \ = \ \rho^2 \cdot t^{}_3 H_{\mu\nu}
\label{Ppsi}
\end{equation}
This is clearly invariant under (\ref{intu1}) and (\ref{intz2}).
In particular, we conclude that the vector field $e_\mu$ determines
a $U_I(1)$ principal bundle over $Gr(4,2)$. 

We employ $H_{\mu\nu}$ to 
explicitely resolve for the $U_I(1)$ structure as follows \cite{fad1}.
We first introduce the electric and magnetic components of (\ref{Ppsi}),
\begin{equation}
\begin{matrix}
E_i = \frac{i}{2} ( e_0^{} e_i^\star - e_i^{} e_0^\star)
\\
B_i = \frac{i}{2} \epsilon_{ijk} e^\star_j e^{}_k
\end{matrix}
\label{EB}
\end{equation}
They are subject to
\begin{equation}
\begin{matrix}
\vec E \cdot \vec B = 0
\\
\vec E \cdot \vec E + \vec B \cdot \vec B = \frac{1}{4}
\end{matrix}
\label{eebb}
\end{equation}
We then define the selfdual and anti-self-dual combinations
\begin{equation}
\vec s_{\pm} = 2 ( \vec B \pm \vec E)
\label{s+-}
\end{equation}
This gives us two independent unit vectors that parametrize the two-spheres
$\mathbb S^2_\pm$ of our Grassmannian $Gr(4,2) \simeq \mathbb S^2_+ \times 
\mathbb S^2_-$, respectively. In these variables
\begin{equation}
e_\mu = \frac{1}{2} e^{i\eta} \cdot 
\left( \sqrt{ 1-\vec s_+ \cdot \vec s_- } \ ,
\ \frac{ \vec s_+ \times \vec s_- + i ( \vec s_- - \vec s_+) }{\sqrt{
1 - \vec s_+ \cdot \vec s_- }} \right) \ = \ e^{i\eta} \cdot 
\left( \sqrt{2 \vec E \cdot \vec E} \ , \ \frac{ 2 \vec E \times 
\vec B - i \vec E } { \sqrt{ 2 \vec E \cdot \vec E} } \right)
\ \equiv \ e^{i\eta} \hat e_\mu^{}
\label{es}
\end{equation}
Here the phase factor $\eta$ describes locally a section of 
the $U_I(1)$ bundle determined by $e_\mu$ 
over the Grassmannian (\ref{Gr}). The $U_I(1)$ transformation 
sends $\eta \to \eta - \lambda$. 

We note that since any two components of $e_\mu$ can vanish
simultaneously, at least three coordinate patches for the base
are needed in order to define the bundle. With local
trivialization determined by $\eta_\alpha = Arg (e_\alpha)$ these patches
can be chosen to be $\mathcal U_\alpha = \{ |e_\alpha| > \epsilon \}$ 
for $\alpha=0,1,2$ with some (infinitesimal)
$\epsilon >0$. On the overlaps $\mathcal U_\alpha  
\bigcap \mathcal U_\beta$ the transition functions are then
\[
f_{\alpha\beta} = \exp\{i \cdot Arg \frac{e_{\beta}}{e_{\alpha}}\}
\]
with $e_{\alpha}$ {\it resp.} $ e_{\beta}$ a component of 
vector $e_\mu$ that is nonvanishing
in the overlap of $\mathcal U_\alpha$ and $\mathcal U_\beta$.

We now proceed to show by explicit computation the nontriviality of 
the $U_I(1)$ bundle. This implies that the phase factor $\eta$ in (\ref{es}) 
can not be globally removed. We do this by relating our $U_I(1)$ bundle to the 
Dirac monopole bundle (Hopf fibration) $\mathbb S^3 \sim
\mathbb S^2 
\times \mathbb S^1 $. We start by introducing the $U_I(1)$ connection 
\begin{equation}
\Gamma = i e_\mu^\star d e_\mu \ = \ i \hat e_\mu^\star d
\hat e_\mu + d\eta \ = \  \ \hat \Gamma + d\eta
\label{Gamma}
\end{equation}
We locally parametrize the vectors $\vec s_\pm$ by
\begin{equation}
\vec s_\pm \ = \ \left( \begin{matrix}
\cos \phi_\pm  \, \sin\theta_\pm \\
\sin \phi_\pm \, \sin\theta\pm \\
\cos \theta_\pm
 \end{matrix} \right) 
\label{paras}
\end{equation}
We substitute this in (\ref{es}), (\ref{Gamma}).
This gives us a (somewhat complicated) expression of $\hat \Gamma$ in terms of the angular
variables (\ref{paras}). When 
we compute the ensuing curvature two-form
the result is
\begin{equation}
F = d\hat \Gamma \equiv d \Gamma = \sin \theta_+ 
d \theta_+ \wedge d \phi_+ + \sin\theta_- d \theta_- 
\wedge d \phi_-
\label{F}
\end{equation}
Consequently the connection $\Gamma$ in (\ref{Gamma}) is gauge equivalent to
a connection of the form
\begin{equation}
\Gamma \sim -\cos \theta_+ d\phi_+ - \cos \theta_- d\phi_- + d\eta
\label{dira1}
\end{equation}
When we restrict to one of the two-spheres $\mathbb S_\pm$ 
in $Gr(4,2)$ by fixing some point ($pt$) in the other, 
we obtain the two submanifolds $\mathbb 
S^2_+ \times pt$ and $pt \times \mathbb S^2_-$ and 
arrive at the functional form of the
Dirac monopole connection in each of them. 
Thus the $U_I(1)$ bundle is non-trivial and admits no global sections, in particular
the section $\eta$ can only be defined locally.

We now proceed to consider the linear action of Euclidean 
(Lorentz) boosts. 
For this we rotate $e_\mu$ to a generic 
spatial direction $\varepsilon_i$ ($i=1,2,3$). 
In the case of an infinitesimal $\varepsilon = 
\sqrt{\vec \varepsilon\cdot \vec
\varepsilon} \ $ the four-vector
$e^{}_\mu$ ($\mu = 0,i)$ 
transforms under the ensuing boost $\Lambda_\varepsilon$
as follows,
\begin{equation}
\begin{array}{c}
\Lambda_{\varepsilon} e^{}_0 = - \varepsilon_i e^{}_i \\
\Lambda_{\varepsilon} e^{}_i = - \varepsilon_i  e^{}_0 .
\end{array}
\label{b1inf}
\end{equation}
For a finite $\varepsilon$ the boost is obtained by
exponentiation,
\begin{equation}
\begin{array}{ccc}
e^{\Lambda_\varepsilon}(e_i)  &=& e_i + 
\frac{1}{\varepsilon^2} \cdot \varepsilon_i ( \vec e \cdot \vec 
\varepsilon \cos (\varepsilon) + \varepsilon e_0 \sin(\varepsilon)
  - \vec e \cdot \vec \varepsilon  )  
\\
e^{\Lambda_\varepsilon}(e_0) &=& e_0 \cos(\varepsilon) - \frac{1}{\varepsilon}
\vec e \cdot \vec \varepsilon \sin(\varepsilon) \ \equiv 
\ 
e_\mu \hat \varepsilon_\mu
\end{array}
\label{b1fin}
\end{equation}
where
\[
0 \le \ \varepsilon \equiv \sqrt{\vec \varepsilon \cdot 
 \vec \varepsilon} \ < 2 \pi \ \ \ \ \ \ (mod \ 2\pi)
\]
and
\[
\hat \varepsilon_\mu = \left( \cos(\varepsilon) \ , \ - 
\sin ( \varepsilon) \frac{\vec \varepsilon}{\varepsilon}\right)
\]

We now identify a different, {\it projective} representation of $SO(4)$ on the
Grassmannian: On the base manifold the ensuing $SO(4)$ 
boost acts on the
electric and magnetic vectors $\vec E$ and $\vec B$ so that 
the result is the familiar
\begin{equation}
\begin{array}{c}
\Lambda_\varepsilon \vec E \equiv \delta_\varepsilon {\vec E} =
\vec B \times \vec \varepsilon \\
\Lambda_\varepsilon \vec B 
\equiv \delta_\varepsilon {\vec B} = \vec E \times \vec \varepsilon.
\end{array}
\label{b2inf}
\end{equation}
For finite boost we get
\begin{equation}
e^{\delta_\varepsilon}(\vec E) = \frac{\vec \varepsilon \, 
(\vec \varepsilon \cdot {\vec E}) 
\, (1-\cos \varepsilon) + [{\vec B} \times \vec \varepsilon ] 
\, \varepsilon \sin \varepsilon + \vec E \; \varepsilon^2 \cos 
\varepsilon }
{\varepsilon^2}
\label{b2fin}
\end{equation}
and the same holds for the finite boost of $\vec B$, but 
with $\vec E$ and  $\vec B$ interchanged.

We assert that the difference between (\ref{b1inf}) and 
(\ref{b2inf}), {\it resp.} (\ref{b1fin}) and (\ref{b2fin}),
is a one-cocycle, due to the projective nature of the second 
representation of $SO(4)$ on $Gr(4,2)$. For this we recall
the definition of a one-cocycle: If $\xi$ denotes a local 
coordinate system on $Gr(4,2)$ and if a section of the $U_I(1)$ 
bundle which is locally specified by $e^{i \eta}$ is 
denoted by $\Psi$, then we have for a projective representation
\begin{equation}
\Lambda(g) \Psi(\xi) = \mathcal C(\xi,g) \Psi(\xi^g)
\label{1coc}
\end{equation}
with $g \in SO(4)$. The factor $\mathcal C(\xi, g)$ is a one-cocycle 
that determines the lifting of the projective representation to
the linear representation.
For a boost with the group element $g \in SO(4)$ 
which is parameterized by (finite) $\vec \varepsilon$ on the base manifold with
$\vec E$ and $ \vec B$, (\ref{1coc}) becomes
\begin{equation}
e^{\Lambda_\varepsilon} \Psi(\vec E , \vec B) = 
\mathcal C(\vec E , \vec B, \vec \varepsilon) 
\Psi\left(e^{\delta_\varepsilon}(\vec E),e^{\delta_\varepsilon}(\vec B)\right)
\label{1coc2}
\end{equation}

We compute the one-cocycle in (\ref{1coc2}) on 
a chart $\mathcal U_0$ with local trivialization 
$\eta = Arg( e_{0})$. With  
$C(\xi,g) = \exp\{i \Theta(\xi,g)\}$ we look at the transformation 
of a local section $\exp\{\eta \}$ under the boost $g$.
Under an infinitesimal boost the phase of $e_0$ changes as
follows \cite{fad1},
\begin{equation}
\Lambda_\varepsilon  \eta = \Theta(\varepsilon) 
= \frac{\vec E \cdot \vec \varepsilon}{2 { \vec E}^2} 
= \frac{(\vec s_+ - \vec s_-) \cdot \vec \varepsilon}{1-\vec 
s_+ \cdot \vec s_-}
\label{1cocycle}
\end{equation}
For a finite boost we find by exponentiation
\begin{equation}
\Theta(\vec \varepsilon) 
\ = \ Arg \left( \hat e_\mu \hat \varepsilon_\mu \right)
\label{cocycle}
\end{equation}
which reduces to (\ref{1cocycle}) for infinitesimal $\epsilon$.
For general $g \in SO(4)$ we get in the chart $\mathcal U_{0}$
\begin{equation}
\Theta(\xi,g) = Arg\left( \frac{e^g_{\;0}}{e^{}_0}\right)
\label{SO4}
\end{equation}

Finally, since all one-dimensional representations are necessarily Abelian
we conclude that $\Theta$ satisfies the one-cocycle condition
\[
\Lambda_{\varepsilon_1} \Theta( {\vec E,
 \vec B }; \vec \varepsilon_2 ) \ - \
\Lambda_{\varepsilon_2} \Theta( {\vec E,  \vec B} ;
 \vec \varepsilon_1 )  \ = \ 0
\]
with its nontriviality following from the nontriviality of the 
Dirac monopole bundles.

\vskip 0.3cm
In conclusion, we have established the nontriviality of the
infinitesimal one-cocycle found in \cite{fad1} by relating it to
the Dirac monopole bundle. We have also reported its finite version.
The presence of the one-cocycle establishes that in spin-charge
separated Yang-Mills theory Lorentz boosts have two inequivalent 
representations, one acting linearly on the Grassmannian $Gr(2,4)$
and the other projectively. The physical consequences of this
observation remain to be clarified; in \cite{fad1} a relation to
Yang-Mills mass gap has been proposed. 

\underline{Acknowledgements}:
We thank Ludvig Faddeev for discussions and comments.
S.S. also thanks David Marsh for discussions. Our work 
has partially been supported by a VR Grant 2006-3376.
The work by S.S. has also been partially supported by the Dmitri 
Zimin 'Dynasty' foundation, RSGSS-1124.2003.2; RFFI project 
grant 06-02-16786, by a STINT Institutional Grant  IG2004-2 025.
A.J.N. acknowledges the hospitality of CERN during the completion
of this work.

\end{document}